\documentclass[pre,reprint,showpacs,amsmath,amssymb,floatfix]{revtex4-1}

\usepackage{graphicx,dcolumn,times}

\def\<{\langle}
\def\>{\rangle}
\newcommand{\rnds}{{\mathfrak s_i}}
\newcommand{\rndstot}{{\mathcal S}}
\newcommand{\bc}{BC}

\newcommand{\NAT}{Nature}
\newcommand{\EPL}{EPL}
\newcommand{\JSTAT}{J.~Stat.~Mech.}

\newcommand{\PRE}{Phys.~Rev. E}

\newcommand{\JPCM}{J.~Phys.: Condens.~Matter}
\newcommand{\JPAOLD}{J.~Phys.~A: Math.~Gen.}

\newcommand{\physrep}{Phys.~Rep.}

\newcommand{\ZPB}{Z.~Phys.~B}

\newcolumntype{.}{D{.}{.}{-1}}

\begin{document}
\title{Critical Casimir force in the presence of random local adsorption preference}

\author{\firstname{Francesco} \surname{Parisen Toldin}}
\email{francesco.parisentoldin@physik.uni-wuerzburg.de}
\affiliation{Institut f\"ur Theoretische Physik und Astrophysik, Universit\"at W\"urzburg, Am Hubland, D-97074 W\"urzburg, Germany}


\pacs{05.70.Jk,68.15.+e,05.50.+q,05.10.Ln}

\begin{abstract}
We study the critical Casimir force for a film geometry in the Ising universality class. We employ a homogeneous adsorption preference on one of the confining surfaces, while the opposing surface exhibits quenched random disorder, leading to a random local adsorption preference. Disorder is characterized by a parameter $p$, which measures, on average, the portion of the surface that prefers one component, so that $p=0$, $1$ correspond to homogeneous adsorption preference. By means of Monte Carlo simulations of an improved Hamiltonian and finite-size scaling analysis, we determine the critical Casimir force. We show that by tuning the disorder parameter $p$, the system exhibits a crossover between an attractive and a repulsive force. At $p=1/2$, disorder allows to effectively realize Dirichlet boundary conditions, which are generically not accessible in classical fluids.
Our results are relevant for the experimental realizations of the critical Casimir force in binary liquid mixtures.
\end{abstract}

\maketitle
\section{Introduction}
\label{sec:intro}
When a fluid close to a critical point is confined between two surfaces, thermal fluctuations give rise to an effective force between the confining surfaces \cite{FG-78}. The resulting fluctuation-induced force is called critical Casimir force and is the analog of the Casimir effect in quantum electrodynamics \cite{Casimir-48}. In recent years, the critical Casimir force has attracted numerous experimental and theoretical investigations, in particular in view of possible technological applications \cite{KG-99,*Ball-07}; see Refs.~\cite{Krech-94,Krech-99,BDT-00,Gambassi-09} for reviews on the topic and also the updated list of references in Ref.~\cite{PTD-10}.
The critical Casimir force is characterized by a universal scaling function, which depends on the universality class (UC) of the bulk phase transition, as well as on the shape of the confinement and on the boundary conditions ({\bc}) therein, through the so-called surface universality classes \cite{Binder-83,Diehl-86}.

Critical Casimir forces have been first indirectly measured by studying the thickness of wetting layers of $^{\rm 4}$He \cite{GC-99,*GSGC-06}, and of classical \cite{FYP-05,*RBM-07} and quantum \cite{GC-02,*UBMCR-03} binary mixtures, close to the critical point. More recently, a direct measure of the critical Casimir force has been obtained by monitoring individual colloidal particles immersed into a critical binary mixture and exposed to a substrate \cite{HHGDB-08,GMHNHBD-09,SZHHB-08,NHB-09,NDHCNVB-11,ZAC-11}. The critical Casimir force has also been studied through its influence on colloidal aggregation \cite{BOSGWS-09,*BOSGWS-09_gd-10,*BOSGWS-09_reply,VAWPMSW-12,NFHWS-13,PMVWMSW-14}.

Until recently theoretical investigations of the critical Casimir force have mainly used field-theoretical methods (see, e.g., Ref.~\cite{PTD-10} for a list of references). Only recently their quantitatively reliable determination has been obtained by means of Monte Carlo (MC) simulations. Early numerical simulations have been used in Ref.~\cite{Krech-97} to compute the critical Casimir force for the three-dimensional film geometry with laterally homogeneous {\bc}. More recently, MC simulations have allowed to determine the critical Casimir force for the $XY$ UC \cite{DK-04,Hucht-07,VGMD-07,VGMD-08,*VGMD-08_erratum,Hasenbusch-09b,Hasenbusch-09c,Hasenbusch-09d} with Dirichlet {\bc}, which describes the critical behavior of the superfluid phase transition in $^{\rm 4}$He, and for the Ising UC \cite{DK-04,VGMD-07,VGMD-08,*VGMD-08_erratum,Hasenbusch-10c,PTD-10,HGS-11,Hasenbusch-11,VMD-11,Hasenbusch-12,Hasenbusch-12b,VED-13,PTTD-13,VD-13,Vasilyev-14,HH-14,LCJH-14,PTTD-14,Hasenbusch-14} with various {\bc}, which describes, {\it inter alia}, the critical behavior of the demixing phase transition in binary liquid mixtures.

For the latter system the order parameter is given by the deviation of the concentration of one of the two species with respect to the critical concentration. In a classical binary liquid mixture the surfaces involved generically prefer one component of the mixture, resulting in an increase of the absolute value of the order parameter close to the surfaces. The force is found to be attractive (respectively, repulsive) when the two surfaces prefer the same (respectively, opposite) component of the mixture. This property, first predicted on the basis of mean-field theory \cite{Krech-97}, has been later confirmed by MC simulations \cite{VGMD-07,VGMD-08,*VGMD-08_erratum,Hasenbusch-10c} and by experimental studies \cite{FYP-05,*RBM-07,HHGDB-08,GMHNHBD-09}. Experimental realizations of the critical Casimir force for colloidal particles immersed into binary liquid mixtures have proven to be very flexible in creating different {\bc} for the surfaces involved. Beside surfaces with a homogeneous adsorption preference \cite{HHGDB-08,GMHNHBD-09}, the critical Casimir force has been investigated in the presence of a chemically structured substrate \cite{SZHHB-08}, leading to a laterally varying adsorption preference, as well as in the presence of a substrate with a gradient in the adsorption preference \cite{NHB-09}. Theoretical investigations of the critical Casimir force for inhomogeneous {\bc} have considered the film geometry in the presence of a chemically striped substrate, studied within mean-field theory \cite{SSD-06}, within Gaussian approximation \cite{KPHSD-04,KPHD-06}, and recently by MC simulations \cite{PTD-10,PTTD-13,PTTD-14}. The critical Casimir force in the presence of a chemically structured substrate has also been studied within the Derjaguin approximation for a sphere \cite{TKGHD-09,*TKGHD-10} and a cylindrical colloid \cite{LLTHD-14} close to a planar wall. Inhomogeneous {\bc} have been also considered within Gaussian approximation in Refs.~\cite{KPHNSP-14,*KPHNSP-14_erratum,KPHNSHP-14,ZRK-04,*ZSRKC-07}.

In this paper, we investigate the critical Casimir force for a system that exhibits quenched random disorder on one of its confining surfaces.
The influence of defects and quenched disorder on surface critical phenomena has attracted much interest; see Ref.~\cite{Pleimling-review} for a recent review. In this context, an important issue is whether disorder is a relevant or irrelevant perturbation. This problem has been extensively investigated in Ref.~\cite{DN-90}, where Harris-type criteria have been formulated. For a semiinfinite three-dimensional system, in agreement with early results on random-field surface disorder \cite{MN-88}, uncorrelated random surface field with null expectation value and uncorrelated random surface couplings are found to be irrelevant at the ordinary transition \cite{DN-90}. For the latter type of disorder the irrelevance on the surface magnetization critical exponent can be rigorously established, provided that the surface bonds do not exceed the threshold of the special transition \cite{Diehl-98}. Nevertheless, a random surface field wipes out the two-dimensional surface phase transition \cite{IM-75}, so that there is no multicritical point associated with the surface special transition.

It is useful to remark that Harris-type criteria are obtained by a perturbative treatment of disorder to the lowest order about the ``pure'' fixed point. Therefore, they are expected to be predictive for weak disorder: even when a Harris-type criterion predicts disorder to be an irrelevant perturbation at a fixed point, sufficiently strong disorder may result in a perturbation that escapes the basin of attraction of the ``pure'' fixed point, leading to a new fixed point for RG flow. An example in this sense is provided by the three-dimensional gauge-glass model, which is an $XY$ model with random phase shifts. For this model, disorder can be considered as a perturbation that effectively couples to the energy \cite{AV-10}. In agreement with the corresponding Harris criterion \cite{Harris-74}, weak disorder is irrelevant and the model exhibits a line of second-order phase transitions in the unperturbed three-dimensional $XY$ UC \cite{AV-10}. Nevertheless, for sufficiently strong disorder the model undergoes a continuous phase transition between a paramagnetic and a glassy phase, with a multicritical point separating the paramagnetic, ferromagnetic, and glassy phases \cite{AV-10}. Another example is provided by the 2D $\pm J$ Ising model. Here, weak disorder is marginally irrelevant, so that for weak disorder the critical behavior is controlled by the ``pure'' 2D Ising fixed point \cite{HPTPV-08b}. Nevertheless, the model exhibits also a strong-disorder fixed point \cite{PHP-06,PTPV-09} and a multicritical point, which in the phase diagram separates the transition line in the Ising UC and the one in the strong-disorder UC \cite{HPTPV-08,PTPV-09}.

In this paper, we study the critical Casimir force for a system in the Ising UC and in the film geometry, such that one confining surface displays a homogeneous adsorption preference while the opposite surface exhibits quenched disorder, leading to a random local adsorption preference. To this end, we combine numerical integration with MC simulations of an improved spin model on a three-dimensional lattice. We introduce a parameter $p$, which controls the fraction of the disordered surface that prefers one component, such that for $p=0$ and $p=1$ we recover a homogeneous adsorption preference, while for $p=1/2$ on average there is no preferential adsorption for one of the components. This setup is equivalent to the presence of an infinitely strong random field on the disordered surface, such that for $p=1/2$ the expectation value of the surface field vanishes. In agreement with the aforementioned Harris-type criterion, we find that in this case the surface effectively realizes a Dirichelet {\bc}, which corresponds to the ordinary UC. For $p\ne 1/2$, we observe a crossover to the limiting cases of homogeneous adsorption preference $p=0$ and $p=1$.

This paper is organized as follows. In Sec.~\ref{sec:model} we introduce the lattice model studied here and summarize the finite-size scaling (FSS) behavior. In Sec.~\ref{sec:method} we discuss the numerical methods used to compute the critical Casimir force. In Sec.~\ref{sec:results} we present our MC results. In Sec.~\ref{sec:summary} we summarize our findings.

\section{Model and Finite-Size Scaling}
\label{sec:model}
We study an improved lattice model \cite{PV-02}, whose critical behavior belongs to the Ising UC. As in recent numerical studies of the critical Casimir force \cite{Hasenbusch-10c,PTD-10,Hasenbusch-11,Hasenbusch-12b,PTTD-13,PTTD-14,Hasenbusch-14}, we consider the Blume-Capel model \cite{Blume-66,Capel-66}, which is defined on a simple cubic lattice, where the spin variables $S_i$ on each site $i$ can take values $+1$, $0$, $-1$. The reduced Hamiltonian ${\cal H}$ is
\begin{equation}
\label{bc}
{\cal H}=-\beta\sum_{<i j>}S_i S_j + D\sum_i S_i^2,\qquad S_i=-1,0,1,
\end{equation}
so that the Gibbs weight is $\exp(-\cal H)$. As done in previous recent investigations of this model \cite{Hasenbusch-10c,PTD-10,Hasenbusch-11,Hasenbusch-12b,PTTD-13,Hasenbusch-10,PTTD-14,Hasenbusch-14}, in the following we shall keep $D$ constant, treating it as a part of the integration measure over the spin configurations, while we vary the parameter $\beta$, which controls the distance to the critical point. In the limit $D\rightarrow -\infty$, the model reduces to the usual Ising model. Starting from $D\rightarrow -\infty$, the phase diagram of the Blume-Capel model displays a line of second-order phase transitions in the Ising UC, which ends at a tricritical point $D_{\rm tri}$. For $D>D_{\rm tri}$, the transition is of first order. In three dimensions, $D_{\rm tri}$ has been determined as $D_{\rm tri}=2.006(8)$ in Ref.~\cite{Deserno-97}, as $D_{\rm tri}\simeq 2.05$ in Ref.~\cite{HB-98}, and more recently as $D_{\rm tri}=2.0313(4)$ in Ref.~\cite{DB-04}. At $D=0.656(20)$ \cite{Hasenbusch-10} the model is improved, i.e., the leading scaling correction $\propto L^{-\omega}$, with $\omega=0.832(6)$ \cite{Hasenbusch-10}, is suppressed. As in recent numerical investigations of the critical Casimir force \cite{Hasenbusch-10c,PTD-10,Hasenbusch-11,Hasenbusch-12b,PTTD-13,PTTD-14,Hasenbusch-14}, here we fixed $D=0.655$. For such a value of $D$, the model is critical for $\beta=\beta_c=0.387721735(25)$ \cite{Hasenbusch-10}.

We consider a three-dimensional film geometry, with lateral extension $L_\parallel$ and thickness $L$, with $L_\parallel\gg L$. We impose periodic {\bc} on the two lateral directions, and fixed {\bc} on the two confining surfaces. On the upper surface we fix the spins to $S_i=1$, mimicking a homogeneous adsorption preference. The spins on the lower surface $\rnds$ are fixed to $\pm 1$ according to the probability distribution
\begin{equation}
{\cal P}\left(\rnds\right) = p\delta\left(\rnds,1\right) + (1-p)\delta\left(\rnds,-1\right).
\label{pdistribution}
\end{equation}
This choice of {\bc} corresponds to a random local adsorption preference where, locally, the substrate prefers one component with probability $p$. For $p=0$ the {\bc} reduce to the so-called $(+,-)$ {\bc}, i.e., to that of a homogeneous strong adsorption preference, with opposite adsorption preference for the two confining surfaces. For $p=1$, we recover the so-called $(+,+)$ {\bc}, where the confining surfaces are characterized by an identical adsorption preference.
In the presence of quenched random disorder, one distinguishes between two averages, the thermal average $\<\ldots\>$, i.e., the average over the Gibbs measure at a given disorder configuration $\{\rnds\}$, and the average $[\ldots]$ over the disorder realizations.

The reduced free-energy density $F(\beta,L,L_\parallel,p)$, i.e, the free energy per volume $V$ and in units of $k_BT$ is given by
\begin{equation}
F(\beta,L,L_\parallel,p)\equiv -\frac{1}{V}\left[\ln\left(\frac{Z(\beta,L,L_\parallel;\{\rnds\})}{Z(\beta=0,L,L_\parallel;\{\rnds\})}\right)\right],
\label{Fdef}
\end{equation}
where $V\equiv LL_\parallel^2$ and $Z(\beta,L,L_\parallel;\{\rnds\})$ is the partition function at a given disorder realization $\{\rnds\}$:
\begin{equation}
Z(\beta,L,L_\parallel;\{\rnds\})=\sum_{\{S_i\}}e^{-\cal H}.
\label{Z}
\end{equation}
The denominator in Eq.~(\ref{Fdef}) fixes the normalization of $F(\beta,L,L_\parallel,p)$ such that $F(\beta=0,L,L_\parallel,p)=0$ \footnote{This choice is equivalent to a shift of the free energy that does not contribute to the critical Casimir force.}. In line with the prescription of quenched random disorder, in Eq.~(\ref{Fdef}) the average over the disorder distribution $[\ldots]$ is done after taking the logarithm of the partition function.
The reduced bulk free-energy density $F_{\rm bulk}(\beta)$ is obtained by taking the thermodynamic limit of $F(\beta,L,L_\parallel,p)$:
\begin{equation}
F_{\rm bulk}(\beta)=\lim_{L,L_\parallel\rightarrow\infty}F(\beta,L,L_\parallel,p).
\label{Fbulk}
\end{equation}
Since $F_{\rm bulk}(\beta)$ is independent of the {\bc}, it does not depend on $p$ either. The reduced excess free-energy density $F_{\rm ex}(\beta,L,L_\parallel,p)$ is defined as the remainder of $F(\beta,L,L_\parallel,p)$ after having subtracted its thermodynamic limit $F_{\rm bulk}(\beta)$
\begin{equation}
F_{\rm ex}(\beta,L,L_\parallel,p) \equiv F(\beta,L,L_\parallel,p)-F_{\rm bulk}(\beta).
\label{Fred}
\end{equation}
The critical Casimir force $F_C$ per area $L_\parallel^2$ and in units of $k_BT$ is defined as
\begin{equation}
\begin{split}
F_C(\beta,L,L_\parallel,p)&\equiv-\frac{\partial \left(LF_{\rm ex}\right)}{\partial L}\Big|_{\beta,L_\parallel,p}\\
&=-\frac{\partial \left(LF\right)}{\partial L}\Big|_{\beta,L_\parallel,p}+F_{\rm bulk}(\beta).
\end{split}
\label{casimir_def}
\end{equation}
Close to a critical point, FSS theory predicts a universal scaling behavior of the critical Casimir force. A general review of FSS theory can be found in Ref.~\cite{Privman-90}, whereas a detailed discussion thereof in the context of critical Casimir forces can be found in Ref.~\cite{PTD-10}.
According to renormalization-group (RG) theory \cite{Wegner-76} and FSS theory \cite{Privman-90}, the leading scaling behavior of $F_C$ can be expressed as
\begin{equation}
\begin{split}
\label{casimir_fss_leading}
F_C(\beta,L,L_\parallel,p)=\frac{1}{L^3}\theta\left(\tau,\ldots\right),\\
\tau\equiv t\left(L/\xi_0^+\right)^{1/\nu}, \quad t\equiv (\beta_c/\beta-1),
\end{split}
\end{equation}
where $\theta(\tau,\ldots)$ is a universal scaling function and $\xi_0^+$ is the nonuniversal amplitude of the correlation length $\xi$ in the high-temperature phase, which fixes the normalization of $\tau$:
\begin{equation}
\label{xi_crit}
\xi(t\rightarrow 0^+)=\xi_0^+|t|^{-\nu}.
\end{equation}
From Ref.~\cite{Hasenbusch-10c} we infer $\xi_{0}^+=0.4145(4)$ for the improved Blume-Capel model studied here. For the critical exponent $\nu$ we use the recent determination $\nu=0.63002(10)$ \cite{Hasenbusch-10}.

Equation (\ref{casimir_fss_leading}) holds in the so-called FSS limit, i.e., the limit of $L\rightarrow\infty$, $t\rightarrow 0$ at fixed $\xi/L$. In Eq.~(\ref{casimir_fss_leading}), the dots $\ldots$ denote the possible dependence of $\theta(\tau,\ldots)$ on additional scaling variables, the presence of which depends on the {\bc}; accordingly, the FSS limit is taken by keeping also such additional scaling variables fixed.
For instance, in principle the universal scaling function $\theta(\tau,\ldots)$ can depend also on the aspect ratio $\rho\equiv L/L_\parallel$.
For the {\bc} and the film geometry $L_\parallel\gg L$ considered here, the dependence of the critical Casimir force on $\rho$ is expected to be negligible.
The MC data presented in the next section support this observation; therefore, for simplicity here and in the following we shall neglect the dependence of the scaling function $\theta(\tau,\ldots)$ on $\rho$. The aspect-ratio dependence of the critical Casimir force has been investigated in Refs.~\cite{PTD-10,PTTD-14} for a film geometry $L_\parallel\gg L$ and in the presence of a chemically stepped substrate, and in Ref.~\cite{HGS-11} for periodic {\bc}.

\section{Simulation methods}
\label{sec:method}
In this section, we summarize the numerical methods used to compute the critical Casimir force. A more detailed discussion of our implementation of the methods can be found in Refs.~\cite{PTD-10,PTTD-13}. The determination of the critical Casimir force proceeds in two steps. We first replace the partial derivative on the right-hand side of Eq.~(\ref{casimir_def}) with a finite difference, computing the free-energy difference $\Delta F(\beta,L,L_\parallel,p)$ per area $L_\parallel^2$, between a film of thickness $L$ and a film of thickness $L-1$
\begin{multline}
\Delta F(\beta,L,L_\parallel,p) \equiv LF(\beta,L,L_\parallel,p)\\
-(L-1)F(\beta,L-1,L_\parallel,p).
\label{DeltaF}
\end{multline}
By using the definition of the critical Casimir force given in Eq.~(\ref{casimir_def}), one finds \cite{PTD-10}
\begin{equation}
F_C\left(\beta,L-\frac{1}{2},L_\parallel,p\right)=-\Delta F(\beta,L,L_\parallel,p)+F_{\rm bulk}(\beta),
\label{force_DeltaF}
\end{equation}
where the choice of computing $F_C$ at the intermediate thickness $L-1/2$ ensures that no additional corrections $\propto L^{-1}$ are generated in the FSS limit \cite{PTD-10}.

Equation (\ref{casimir_fss_leading}) describes only the leading scaling behavior of $F_C$. In order to extract the universal scaling function $\theta(\tau,\ldots)$ from numerical simulations, it is important to take into account corrections to scaling.
In a finite size, and for the improved lattice model considered here, the amplitude of the leading bulk irrelevant operator is suppressed, so that the leading scaling correction is expected to be due to the presence of nonfully periodic {\bc}. Any {\bc} that are not fully (anti)periodic or, more generally, the absence of translational invariance gives rise to scaling corrections $\propto 1/L$, which, as proposed for the first time in Ref.~\cite{CF-76}, can be absorbed by the substitution $L\rightarrow L+c$, where $c$ is a nonuniversal, temperature--independent length. Recently, this property has been confirmed in several numerical simulations of classical models \cite{Hasenbusch-08,Hasenbusch-09,Hasenbusch-09b,Hasenbusch-10c,PTD-10,Hasenbusch-11,Hasenbusch-12,Hasenbusch-12b,PTTD-13,PTTD-14,Hasenbusch-14} and has been argued to hold also for FSS behavior at a quantum phase transition \cite{CPV-14}. In the present case, upon employing the substitution $L\rightarrow L+c$ in Eq.~(\ref{casimir_fss_leading}) and using it in Eq.~(\ref{force_DeltaF}), we obtain the following FSS Ansatz for $\Delta F(\beta,L,L_\parallel,p)$:
\begin{multline}
\Delta F(\beta,L,L_\parallel,p)= F_{\rm bulk}(\beta)\\
-\frac{1}{(L-1/2+c)^3}\theta\left(t\left(\frac{L-1/2+c}{\xi_0^+}\right)^{1/\nu},\ldots\right).
\label{casimir_fss}
\end{multline}
A detailed discussion on the type of scaling corrections and possible modifications to Eq.~(\ref{casimir_fss}) can be found in Ref.~\cite{PTTD-13}.

Two methods for computing $\Delta F(\beta,L,L_\parallel,p)$ have gained popularity in recent numerical investigations. In the coupling parameter approach introduced in Ref.~\cite{VGMD-07} and also used in Refs.~\cite{VGMD-08,*VGMD-08_erratum,PTD-10,VMD-11,PTTD-13,VED-13,VD-13,Vasilyev-14,PTTD-14}, one defines a crossover Hamiltonian ${\cal H}_\lambda\equiv \left(1-\lambda\right){\cal H}_1 + \lambda{\cal H}_2$, which depends on a parameter $\lambda\in [0,1]$ and is a convex combination of the Hamiltonian ${\cal H}_1$ for a film of thicknesses $L$ and of the Hamiltonian ${\cal H}_2$ which corresponds to a film thickness $L-1$. Then $\Delta F(\beta,L,L_\parallel,p)$ is obtained by a numerical integration over $\lambda$ of the thermal average of a suitable observable $\<{\cal H}_2-{\cal H}_1\>_\lambda$ in the Gibbs ensemble described by ${\cal H}_\lambda$. This observable can be computed by standard MC simulations.
An alternative approach, introduced in Ref.~\cite{Hucht-07} and also employed in Refs.~\cite{Hasenbusch-09b,Hasenbusch-09c,Hasenbusch-09d,Hasenbusch-10c,HGS-11,Hasenbusch-11,PTTD-13,HH-14,PTTD-14} consists in evaluating $F(\beta,L,L_\parallel,p)$ for thicknesses $L$ and $L-1$ through a numerical integration over $\beta$, where the integrand $\partial F/\partial\beta$ can be determined by standard MC simulations, and subsequently calculating the free-energy difference in Eq.~(\ref{DeltaF}). Another recent method for computing the Casimir force is the geometric cluster algorithm, which has been employed in Refs.~\cite{Hasenbusch-12b,Hasenbusch-14}.

Finally, the universal scaling function $\theta(\tau,\ldots)$ of the critical Casimir force is obtained by inverting Eq.~(\ref{casimir_fss}). For this purpose, one needs the value of the nonuniversal constant $c$, which can be conveniently determined by studying the Casimir force at criticality, and a determination of the bulk free-energy density $F_{\rm bulk}(\beta)$. For the subtraction of $F_{\rm bulk}(\beta)$, we have used the data obtained in Ref.~\cite{PTTD-13}, where we have determined $F_{\rm bulk}(\beta)$ for the present model, using periodic {\bc} and achieving a precision of $10^{-7}$.

In this work we have also computed the derivative of the critical Casimir force with respect to the disorder parameter $p$. Using Eq.~(\ref{force_DeltaF}), $\partial F_C/\partial p$ can be expressed as
\begin{equation}
\frac{\partial F_C\left(\beta,L-\frac{1}{2},L_\parallel,p\right)}{\partial p}=-\frac{\partial \Delta F(\beta,L,L_\parallel,p)}{\partial p},
\label{derivative_DeltaF}
\end{equation}
where $F_{\rm bulk}(\beta_c)$ drops out because it is disorder-independent. Hence, the determination of $\partial F_C/\partial p$ reduces to the computation of the derivative of the free-energy difference with respect to $p$ and no subsequent subtraction of the bulk free-energy density is needed. Both the coupling parameter approach and the method of thermodynamical integration over $\beta$ rely on a numerical integration of an observable sampled by standard MC simulations. Since the numerical quadrature commutes with $\partial/\partial p$, the evaluation of $\partial \Delta F/\partial p$ reduces to the computation of the derivative with respect to $p$ of the observable used to determine the critical Casimir force. To this end, let us consider in full generality the problem of determining $\partial [\<O\>]/\partial p$, where $O$ is an arbitrary observable. Its mean value is given by
\begin{equation}
\label{meanO}
[\<O\>]=\sum_{\{\rnds=\pm 1\}}\left(\prod_i{\cal P}\left(\rnds\right)\right)\<O\>,
\end{equation}
where the thermal expectation value $\<O\>$ depends implicitly on the disorder realization $\{\rnds\}$. In order to compute $\partial [\<O\>]/\partial p$, we cast ${\cal P}\left(\rnds\right)$ in the form \cite{Nishimori-book}
\begin{equation}
{\cal P}\left(\rnds\right) = \frac{e^{K_p\rnds}}{2\cosh K_p}, \qquad e^{2K_p}=\frac{p}{1-p},
\label{pdistribution_nishimori}
\end{equation}
so that Eq.~(\ref{meanO}) can be expressed as
\begin{equation}
\label{meanO_nishimori}
[\<O\>]=\sum_{\{\rnds=\pm 1\}}\frac{e^{K_p\sum_i\rnds}}{\left(2\cosh K_p\right)^M}\<O\>,
\end{equation}
where $M$ is the number of quenched spins $\rnds$; $M=L_\parallel^2$ in the present case. Using Eq.~(\ref{meanO_nishimori}) it is straightforward to write an expression for $\partial [\<O\>]/\partial p$:
\begin{equation}
\frac{\partial[\<O\>]}{\partial p}=\frac{dK_p}{dp}\left([\rndstot\<O\>]-[\rndstot][\<O\>]\right),\qquad \rndstot\equiv\sum_i\rnds.
\label{derivative}
\end{equation}
Combining the coupling parameter approach with Eq.~(\ref{derivative}), we have determined $\partial\Delta F/\partial p$ for $p=1/2$; our results are presented in Sec.~\ref{sec:crossover}. The evaluation of $\partial\Delta F/\partial p$ turned out to be computationally rather involved because of the statistical noise in the estimator of Eq.~(\ref{derivative}), so that a large number of disorder samples are needed in order to achieve a satisfactory precision. We observe that for $p=1/2$, the second term on the right-hand side of Eq.~(\ref{derivative}) vanishes because $[\rndstot]=0$, so that in principle it can be omitted in computing $\partial [\<O\>]/\partial p$. Nevertheless, in our numerical results at $p=1/2$, we have found necessary to subtract the second term on the right-hand side of Eq.~(\ref{derivative}). In fact, upon substituting the exact disorder average $[\ldots]$ with an average over a {\it finite} number of disorder samples, the large statistical covariance between the first and the second term on the right-hand side of Eq.~(\ref{derivative}) gives rise to a large reduction of the statistical error bar of $\partial [\<O\>]/\partial p$ when the second term is included: without its subtraction, the resulting error bar of $\partial [\<O\>]/\partial p$ largely dominates over its average value, rendering the MC estimates useless.

In numerical simulations of models in the presence of quenched disorder, the sampled observables are computed by averaging the MC data over a finite number of measures $N_{\rm MC}$ at a given realization of the quenched variables, and subsequently averaged over a finite number $N_s$ of disorder samples. The statistical error bars are essentially determined by the number of disorder samples. More precisely, as proven in Appendix B of Ref.~\cite{HPTPV-07}, for most of the observables (including the ones used here to compute the critical Casimir force) the MC sampled average converges in probability to the true value for $N_s\rightarrow\infty$ at fixed $N_{\rm MC}$, irrespective of value of $N_{\rm MC}$. In fact, provided that the MC data are correctly thermalized, one could even sample a single measure $N_{\rm MC}=1$ per sample. In practice, it is convenient to take a sufficiently large value of $N_{\rm MC}$ in order to check the thermalization. We mention that it is possible to optimize the MC run in order to minimize the variance at a fixed computational time; see Appendix B of Ref.~\cite{HPTPV-07} for a discussion of this point.

\section{Results}
\label{sec:results}
\subsection{Critical Casimir force at $p=1/2$}
\label{sec:phalf}
In a series of MC simulations we have computed $\Delta F(\beta=\beta_c=0.387721735,L,L_\parallel,p)$ employing the coupling parameter approach mentioned in Sec.~\ref{sec:method} for film thicknesses $L=8$, $12$, $16$, $24$, $32$, $48$, and $64$, and for three different aspect ratios $\rho=L/L_\parallel=1/8$, $1/12$, and $1/16$. We have sampled the distribution of the randomly frozen spins given in Eq.~(\ref{pdistribution}) with $p=1/2$ by averaging over $N_s$ disorder samples. For lattice sizes $L=48$, $L=64$ we have considered $N_s=100k$ ($1k=10^3$) for $\rho=1/8$, $N_s=50k$ for $\rho=1/12$, and $N_s=25k$ for $\rho=1/16$. For $L=24$, $32$ we have sampled the disorder distribution with $N_s=20k$ for $\rho=1/8$, $N_s=10k$ for $\rho=1/12$, and $N_s=5k$ for $\rho=1/16$. A smaller number of disorder realizations has been considered for smaller lattices. Within the statistical precision our results are independent of the aspect ratio $\rho$; thus, we consider them as reliably describing the limit $\rho\rightarrow 0$.

\begingroup
\squeezetable
\begin{table}
\caption{Fits to Eq.~(\ref{fit_criticality}) for $p=1/2$. $L_{\rm min}$ is the minimum lattice thickness taken into account. $\rm DOF$ denotes the degrees of freedom.}
\label{fit_results_phalf}
\begin{ruledtabular}
\begin{tabular}{l@{}.@{}.@{}.@{}.}
\multicolumn{1}{c}{$L_{\rm min}$} & \multicolumn{1}{c}{$F_{\rm bulk}(\beta_c)$} & \multicolumn{1}{c}{$\Theta$} & \multicolumn{1}{c}{$c$} & \multicolumn{1}{r}{$\chi^2/{\rm DOF}$} \\
\hline
$8$ & -0.07573681(3) & 0.529(3) & -0.90(1) & 15.0/18 \\
$12$ & -0.07573684(4) & 0.523(6) & -0.96(4) & 11.8/15 \\
$16$ & -0.07573689(5) & 0.507(9) & -1.2(1)  & 5.2/12 \\
$24$ & -0.07573689(8) & 0.51(2)  & -1.2(3)  & 3.8/9\\
\end{tabular}
\end{ruledtabular}
\end{table}
\endgroup
By setting $\beta=\beta_c$ in Eq.~(\ref{casimir_fss}), we obtain the expected leading FSS behavior of $\Delta F(\beta=\beta_c,L,L_\parallel,p)$:
\begin{equation}
\Delta F(\beta_c,L,L_\parallel,p)= F_{\rm bulk}(\beta_c)-\frac{\Theta}{(L-1/2+c)^3},
\label{fit_criticality}
\end{equation}
where $\Theta\equiv\theta(\tau=0,\ldots)$ is the Casimir force amplitude at criticality. Here, we are implicitly taking the FSS limit at fixed $p=1/2$.

In Table \ref{fit_results_phalf} we report the results of the fits of $\Delta F(\beta_c,L,L_\parallel,p)$ to Eq.~(\ref{fit_criticality}), leaving $F_{\rm bulk}(\beta_c)$, $\Theta$, and $c$ as free parameters, as a function of the smallest thickness $L_{\rm min}$ used in the fits.
Inspection of the fit results reveals a good $\chi^2/{\rm DOF}$ ($\rm DOF$ denotes the degrees of freedom); however, the fitted values of $\Theta$ display a small dependence on $L_{\rm min}$, which is of the size of the statistical error bars. In order to assess the size of possible subleading or competing scaling corrections, we have considered an alternative Ansatz for the FSS behavior of the critical Casimir force. In the presence of a generic correction-to-scaling term $\propto L^{-\omega}$, Eq.~(\ref{fit_criticality}) should be generalized to \footnote{In general, one may observe a superposition of scaling corrections; in Eq.~(\ref{fit_criticality_omega}) the term $\propto L^{-\omega}$ represents the leading correction-to-scaling term.}
\begin{equation}
\Delta F(\beta_c,L,L_\parallel,p) = F_{\rm bulk}(\beta_c)- \frac{\Theta\cdot\left(1-C(L-1/2)^{-\omega}\right)}{(L-1/2)^3}.
\label{fit_criticality_omega}
\end{equation}
We have fitted our MC data for $\Delta F(\beta_c,L,L_\parallel,p)$ to Eq.~(\ref{fit_criticality_omega}), leaving $F_{\rm bulk}(\beta_c)$, $\Theta$, $C$, and $\omega$ as free parameters. Using all the available lattice sizes we obtain $\Theta=0.52(2)$ and $\omega=1.1(1)$, which is consistent with Eq.~(\ref{fit_criticality}), where the leading scaling correction is $\propto L^{-1}$. Nevertheless, we cannot exclude the presence of additional irrelevant operators with an RG dimension close to $-1$.
Inspecting the fit results reported in Table \ref{fit_results_phalf}, we can take as a conservative estimate of $\Theta$ the fit result for $L_{\rm min}=24$, $\Theta=0.51(2)$, which also agrees with the other results for $L_{\rm min}<24$, including a variation of one error bar, and with the fit to Eq.~(\ref{fit_criticality_omega}). As a further check of this result, we observe that the fitted values of $F_{\rm bulk}(\beta_c)$ fully agree with the result reported in Ref.~\cite{Hasenbusch-10c} $F_{\rm bulk}(\beta_c)=-0.0757368(4)$ and with the values obtained in Ref.~\cite{PTTD-13} for the various {\bc} considered therein. Our result for $\Theta$ can be compared with the critical Casimir force amplitude for a system in the film geometry and in the Ising UC for the so-called $(+,o)$ {\bc}, where one of the confining surfaces exhibits a strong uniform adsorption preference, while the opposite surface has open {\bc}, thus realizing Dirichlet {\bc}, which correspond to the ordinary surface UC. Recent numerical studies have reported $\Theta_{(+,o)}=0.497(3)$ \cite{Hasenbusch-11} and $\Theta_{(+,o)}=0.492(5)$ \cite{PTTD-13}. These values are in full agreement with our result for $p=1/2$ and thus suggest that the resulting critical Casimir force is equivalent to the force for a system where the disordered surface is substituted by a surface where the spin variables are fixed to their expectation value $[\rnds]=2p-1=0$, realizing open {\bc}. Dirichlet {\bc} are generically not accessible in fluids; such {\bc} can also be obtained with a chemically striped surface, in the limit of narrow stripes \cite{PTTD-13}.

\begin{figure}[b]
\includegraphics[width=\linewidth,keepaspectratio]{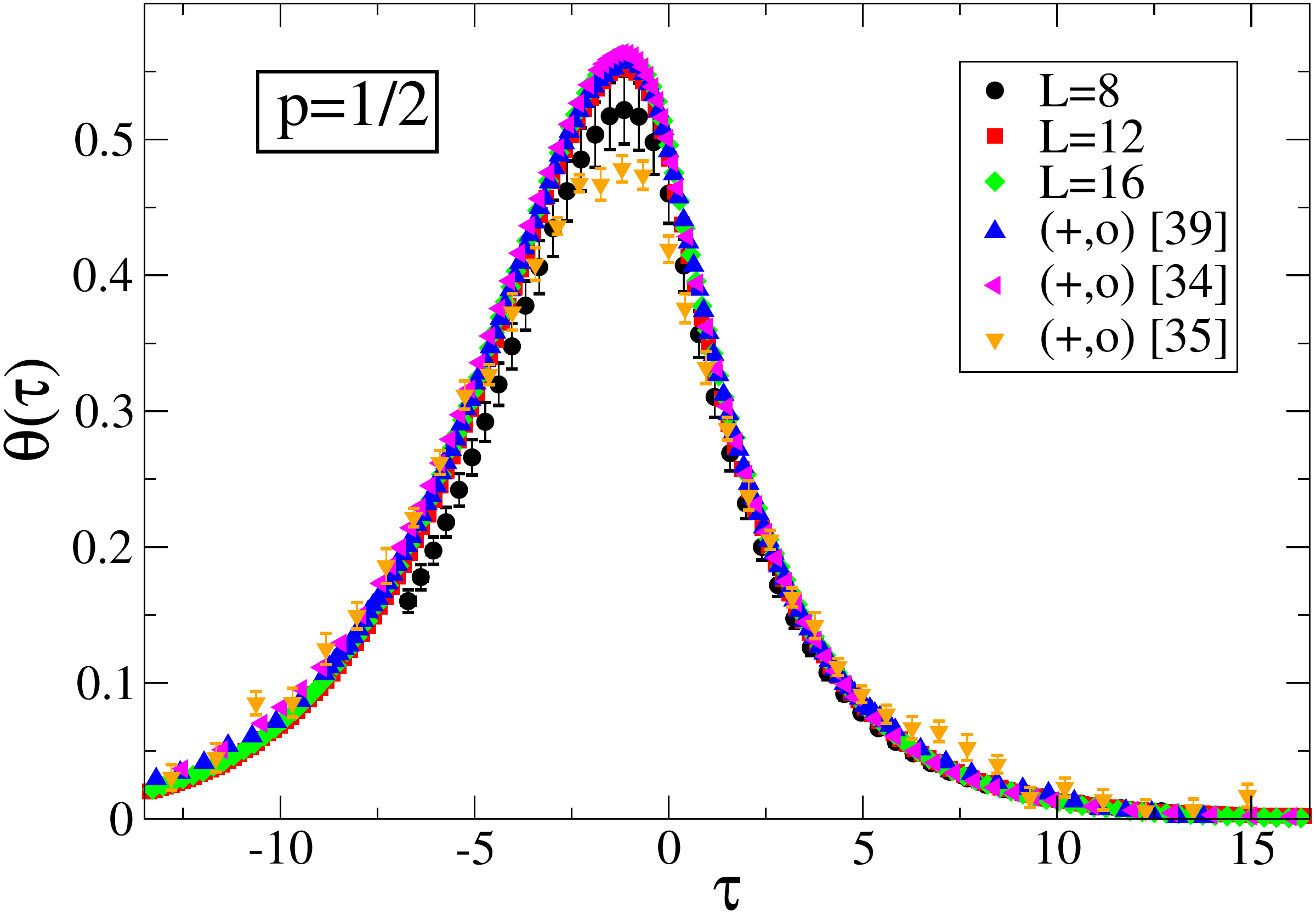}
\caption{(Color online) The universal scaling function $\theta(\tau)$ for $p=1/2$, obtained using $c=-1.2(1)$ as inferred from Table \ref{fit_results_phalf}. We compare our results with the previous determinations of $\theta_{(+,o)}(\tau)$, as obtained in Ref.~\cite{PTTD-13} for $L=24$, in Ref.~\cite{Hasenbusch-11} for $L=16$, and in Ref.~\cite{VMD-11} for $L=20$. The error bars for $L=8$ are obtained as a sum of the statistical error bars stemming from the MC simulations and of the uncertainty in $c$; the latter gives the dominant contribution. The omitted error bars are comparable with the symbol size.}
\label{thetap0.5}
\end{figure}
The disorder setup for $p=1/2$ is equivalent to a random surface field with a bimodal distribution $\pm h$ of vanishing expectation value and in the limit of infinite amplitude $h\rightarrow\infty$. According to a Harris-type criterion for weak surface disorder \cite{DN-90}, random surface field with null expectation value is an irrelevant perturbation at the ordinary transition. Thus, our MC results support the validity of this Harris-type criterion in the limit of infinite amplitude of the random field. This finding is further confirmed by the determination of the critical Casimir scaling function $\theta(\tau)$ for $p=1/2$, which is shown in Fig.~\ref{thetap0.5}. It has been computed using the scheme of thermodynamical integration over $\beta$ (see Sec.~\ref{sec:method}), for $L=8$, $12$, $16$, aspect ratios $\rho=1/8$, $1/12$, $1/16$, and averaging over $N_s=100-1000$ disorder samples. We have checked that within the numerical accuracy our results are independent of the aspect ratio $\rho$ and thus are a reliable extrapolation of the $\rho\rightarrow 0$ limit; in the curves shown in Fig.~\ref{thetap0.5} we have taken the average over the three aspect ratios. In Fig.~\ref{thetap0.5} we also compare our results with those obtained for $(+,o)$ {\bc} in Refs.~\cite{Hasenbusch-11,VMD-11,PTTD-13}. Our results for $L\ge 12$ agree very well with the curves for $\theta_{(+,o)}(\tau)$ as determined in Refs.~\cite{Hasenbusch-11,PTTD-13}. As mentioned in Ref.~\cite{PTTD-13}, the small, but significant, deviation from the curve of Ref.~\cite{VMD-11} may be due to residual scaling corrections in the data of Ref.~\cite{VMD-11}.

\subsection{Crossover behavior for $p\ne 1/2$}
\label{sec:crossover}
Using the coupling parameter approach we have computed $\Delta F(\beta=\beta_c=0.387721735,L,L_\parallel,p)$ for $L=8$, $12$, $16$, $24$, $32$, $48$, and $64$ and for three different aspect ratios $\rho=L/L_\parallel=1/8$, $1/12$, and $1/16$. We have sampled the distribution of the randomly frozen spins given in Eq.~(\ref{pdistribution}) with $p=0.2$, $0.3$, $0.7$, and $0.8$ by averaging over $N_s$ disorder samples. For lattice size $L=64$ we have considered $N_s=100k$ for $\rho=1/8$, $N_s=50k$ for $\rho=1/12$, and $N_s=25k$ for $\rho=1/16$. For $L=48$ we have sampled the disorder distribution with $N_s=50k$ for $\rho=1/8$, $N_s=25k$ for $\rho=1/12$, and $N_s=10k$ for $\rho=1/16$. For $L=24$, $32$ we have considered $N_s=20k$ for $\rho=1/8$, $N_s=10k$ for $\rho=1/12$, and $N_s=5k$ for $\rho=1/16$. A smaller number of disorder realizations has been sampled for smaller lattices. As in the case of the symmetric distribution $p=1/2$, within the statistical precision our results are independent of the aspect ratio $\rho$ and thus they can be regarded as reliably describing the limit $\rho\rightarrow 0$.

\begin{table*}
\caption{Same as Table \ref{fit_results_phalf} for $p=0.2$, $0.3$, $0.7$, $0.8$.}
\label{fit_results}
\begin{ruledtabular}
\begin{tabular}{l|l@{}.@{}.@{}.@{}.}
&\multicolumn{1}{c}{$L_{\rm min}$} & \multicolumn{1}{c}{$F_{\rm bulk}(\beta_c)$} & \multicolumn{1}{c}{$\Theta$} & \multicolumn{1}{c}{$c$} & \multicolumn{1}{c}{$\chi^2/DOF$} \\
\hline
        & $8$           & -0.07573759(3) & 5.378(4) & 0.864(2) & 1030.0/18 \\
$p=0.2$ & $12$          & -0.07573706(3) & 5.533(6) & 1.029(6) & 41.4/15 \\
        & $16$          & -0.07573692(4) & 5.584(11)& 1.10(1)  & 5.8/12 \\
        & $24$          & -0.07573687(7) & 5.61(3)  & 1.13(4)  & 2.3/9 \\
\hline
        & $8$           & -0.07573933(3) & 4.650(5) & 1.478(4) & 6285.3/18 \\
$p=0.3$ & $12$          & -0.07573795(4) & 5.104(8) & 2.075(9) & 390.2/15 \\
        & $16$          & -0.07573742(5) & 5.32(1)  & 2.41(2)  & 39.8/12 \\
        & $24$          & -0.07573705(9) & 5.49(3)  & 2.69(5)  & 5.2/9 \\
\hline
        & $8$           & -0.07573728(4) & -0.946(9) & 4.05(5)  & 137.8/18 \\
$p=0.7$ & $12$          & -0.07573699(5) & -0.83(1)  & 3.13(9)  & 14.3/15 \\
        & $16$          & -0.07573693(6) & -0.81(2)  & 2.8(2)   & 10.7/12 \\
        & $24$          & -0.0757368(1)  & -0.77(4)  & 2.4(5)   & 5.6/9 \\
\hline
        & $8$           & -0.07573689(4) & -0.834(6)& 1.21(2)  & 13.6/18 \\
$p=0.8$ & $12$          & -0.07573689(4) & -0.834(9)& 1.21(6)  & 13.4/15 \\
        & $16$          & -0.07573686(6) & -0.82(2) & 1.1(1)   & 10.8/12 \\
        & $24$          & -0.0757368(1)  & -0.81(4) & 1.0(4)   & 8.6/9 \\
\end{tabular}
\end{ruledtabular}
\end{table*}
For values of $p\ne 1/2$ considered here, the frozen spins $\{\rnds\}$ acquire a nonzero expectation value; therefore, disorder realizes symmetry-breaking {\bc}. Without making any assumption {\it a priori} on the fixed points that control the critical behavior, we have attempted to fit the MC results to Eq.~(\ref{fit_criticality}), leaving $F_{\rm bulk}(\beta_c)$, $\Theta$, and $c$ as free parameters. Such an Ansatz, where no additional scaling variables for $\Theta$ are introduced, corresponds to taking the FSS limit at a fixed value of $p$. Inspection of the fit results reported in Table \ref{fit_results} shows that $\Theta$ changes sign with $p$: the force is attractive for $p>1/2$ and repulsive for $p<1/2$. Furthermore, the fit results for $p<1/2$ (respectively, for $p > 1/2$) appear to approach a common value $\Theta\approx 5.5-5.6$ (respectively, $\Theta\approx -0.8$). For $p< 1/2$, upon increasing $L_{\rm min}$ the fitted value of $\Theta$ approaches the critical Casimir force amplitude for $(+,-)$ {\bc} $\Theta_{(+,-)}=\Theta(p=0)=5.613(20)$~\cite{Hasenbusch-10c}. Indeed, fit results for $p=0.2$ exhibit a small drift as $L_{\rm min}$ is increased. For $L_{\rm min}\ge 12$ the fitted values of $F_{\rm bulk}(\beta_c)$ agree with the previous determination $F_{\rm bulk}(\beta_c)=-0.0757368(4)$ \cite{Hasenbusch-10c} and for $L_{\rm min}=24$, the fitted value of $\Theta=5.61(3)$ matches $\Theta_{(+,-)}=\Theta(p=0)=5.613(20)$~\cite{Hasenbusch-10c}. For $p=0.3$ the fitted values of $\Theta$ exhibit a systematic drift as $L_{\rm min}$ is increased. Nevertheless, the results are compatible with a slow approach to the $(+,-)$ fixed point. In fact, the resulting $\Theta(p=0.3)$ for $L_{\rm min}=24$ differs only by $2\%$ from $\Theta_{(+,-)}=\Theta(p=0)=5.613(20)$~\cite{Hasenbusch-10c}. The fit results for $p=0.7$, $0.8$ are more stable upon increasing $L_{\rm min}$, and $F_{\rm bulk}(\beta_c)$ agrees with the available value $F_{\rm bulk}(\beta_c)=-0.0757368(4)$ \cite{Hasenbusch-10c}. Except for $p=0.7$, $L_{\rm min}=8$, the fitted values of $\Theta$ values agree with the critical Casimir force amplitude for $(+,+)$ {\bc} $\Theta_{(+,+)}=\Theta(p=1)=-0.820(15)$~\cite{Hasenbusch-10c}.

The fit results show that the critical behavior for $p<1/2$ (respectively, $p>1/2$) is controlled by the $p=0$ (respectively, $p=1$) fixed point. Indeed, the present disorder setup where the spins of the lower substrate are fixed according to the probability distribution of Eq.~(\ref{pdistribution}) is equivalent to a film of thickness $L-1$, where the lower spins are subject to an uncorrelated random field with bimodal distribution $\pm\beta$ of nonzero mean value $\beta[\rnds]=\beta(2p-1)$ and standard deviation $2\beta\sqrt{p(1-p)}$. Therefore, disorder can be regarded as a perturbation on a system where the lower confining surface is subject to a finite field $h=\beta(2p-1)$, with $h<0$ for $p<1/2$ and $h>0$ for $p>1/2$. In the presence of a nonzero surface field $h$, the fixed point of the RG flow gives the so-called normal UC \cite{Binder-83,Diehl-86}, which is equivalent to the extraordinary UC \cite{BD-94}; for the system considered here, due to the presence of an opposing surface with spins fixed to $+1$, this corresponds to $(+,-)$ {\bc} for $h<0$ and to $(+,+)$ {\bc} for $h>0$. According to a Harris-type criterion for surface disorder, an uncorrelated random surface field is an irrelevant perturbation at the normal-extraordinary fixed point \cite{DN-90}. Our MC results confirm the validity of this Harris-type criterion for the disorder distribution given in Eq.~(\ref{pdistribution}). We notice that for $p>(2-\sqrt{2})/4\simeq 0.15$ and $p<(2+\sqrt{2})/2\simeq 0.85$ the standard deviation of the probability distribution exceeds its mean value; thus, for the values of $p$ considered here disorder represents a strong perturbation.

Our results clearly show that $p-1/2$ is a relevant perturbation to the $p=1/2$ fixed point. For $p\rightarrow 1/2$ the universal scaling function for the critical Casimir force is expected to acquire an additional scaling variable $(p-1/2)L^y$ and to show a crossover behavior from the $p=1/2$ fixed point to the $p=0$, $1$ fixed points. Therefore, for $p\rightarrow 1/2$ the critical Casimir force is expected to exhibit the following FSS behavior
\begin{equation}
F_C(\beta,L,L_\parallel,p)=\frac{1}{L^3}\theta\left(\tau,\varphi=(p-1/2)L^y\right), \quad p\rightarrow 1/2,
\label{crossover}
\end{equation}
where $y>0$ is the RG dimension of the relevant perturbation and scaling corrections have been neglected. The scaling function $\theta$ introduced in Eq.~(\ref{crossover}) is universal up to a nonuniversal normalization constant of the relevant scaling field $u_p\propto p-1/2$. We note that Eq.~(\ref{crossover}) is expected to be valid in the vicinity of the fixed point $p=1/2$ only; additional corrections stem from higher-order terms in the expansion of the scaling field $u_p\propto(p-1/2) + O(p-1/2)^2$ \cite{AF-83}. Moreover, Eq.~(\ref{crossover}) breaks down at $p=0$, $1$, where the confining surfaces exhibit homogeneous adsorption preference and no additional scaling variables are present \cite{Krech-97,VGMD-07,VGMD-08,*VGMD-08_erratum,Hasenbusch-10c}. For $1/2<p<1$ (respectively, $0<p<1/2$) and a {\it finite} lattice size $L$, the system exhibits a crossover behavior from the $p=1/2$ fixed point to the $p=1$ (respectively, $p=0$) fixed point, which is more evident for values of $p$ closer to $p=1/2$ and for smaller lattice sizes. This is consistent with the fit results of Table \ref{fit_results}, which 
for $p=0.3$ (respectively, for $p=0.7$) show a larger deviation from the previous determination of $\Theta_{(+,-)}$ (respectively, $\Theta_{(+,+)}$) with respect to the corresponding results for $p=0.2$ (respectively, $p=0.8$).

The observed crossover behavior is analogous to the crossover effect on the critical Casimir force due to finite surface fields \cite{Hasenbusch-11,VMD-11,MMD-10}. In view of this analogy, one may expect to identify the RG dimension $y$ in Eq.~(\ref{crossover}) with the RG dimension of the surface field at the ordinary transition $y_{h_1}=0.7249(6)$ \cite{Hasenbusch-11}. The exponent $y$ of Eq.~(\ref{crossover}) can be extracted by computing $\partial\Delta F(\beta_c,L,L_\parallel,p)/\partial p$. By using Eq.~(\ref{crossover}) in Eq.~(\ref{derivative_DeltaF}) and assuming corrections to scaling analog to Eq.~(\ref{casimir_fss}), we find
\begin{equation}
\frac{\partial \Delta F(\beta_c,L,L_\parallel,p)}{\partial p}\Big|_{p=1/2}=A\left(L-1/2+c\right)^{y-3},
\label{derivative_ansatz}
\end{equation}
where $A$ is a nonuniversal constant.
Using the method discussed in Sec.~\ref{sec:method}, we have computed $\partial\Delta F/\partial p$ for $p=1/2$ and lattice sizes $L=8$, $12$, $16$, averaging over $N_s=100k$ samples, and $L=24$ averaging over $N_s=400k$ samples. We have chosen an aspect ratio $\rho=1/8$, which, according to the above results, reliably describes the $\rho\rightarrow 0$ limit. A fit of the MC data to Eq.~(\ref{derivative_ansatz}) leaving $A$ and $y$ as free parameters and setting $c=-1.2(1)$ as inferred from Table \ref{fit_results_phalf} gives $y=0.79(2+3)$, where the first error is the statistical error bar of the fit and the second one reflects the uncertainty in $c$; disregarding the data at $L=8$ we obtain $y=0.68(7+2)$.
Taking into account the fact that, according to the analysis of Sec.~\ref{sec:phalf}, data at $L=8$ are affected by residual scaling corrections, our results support, albeit with a limited precision, the identification of $y$ with the RG dimension of the surface field at the ordinary UC $y_{h_1}=0.7249(6)$ \cite{Hasenbusch-11}.

\begin{figure}[b]
\includegraphics[width=\linewidth,keepaspectratio]{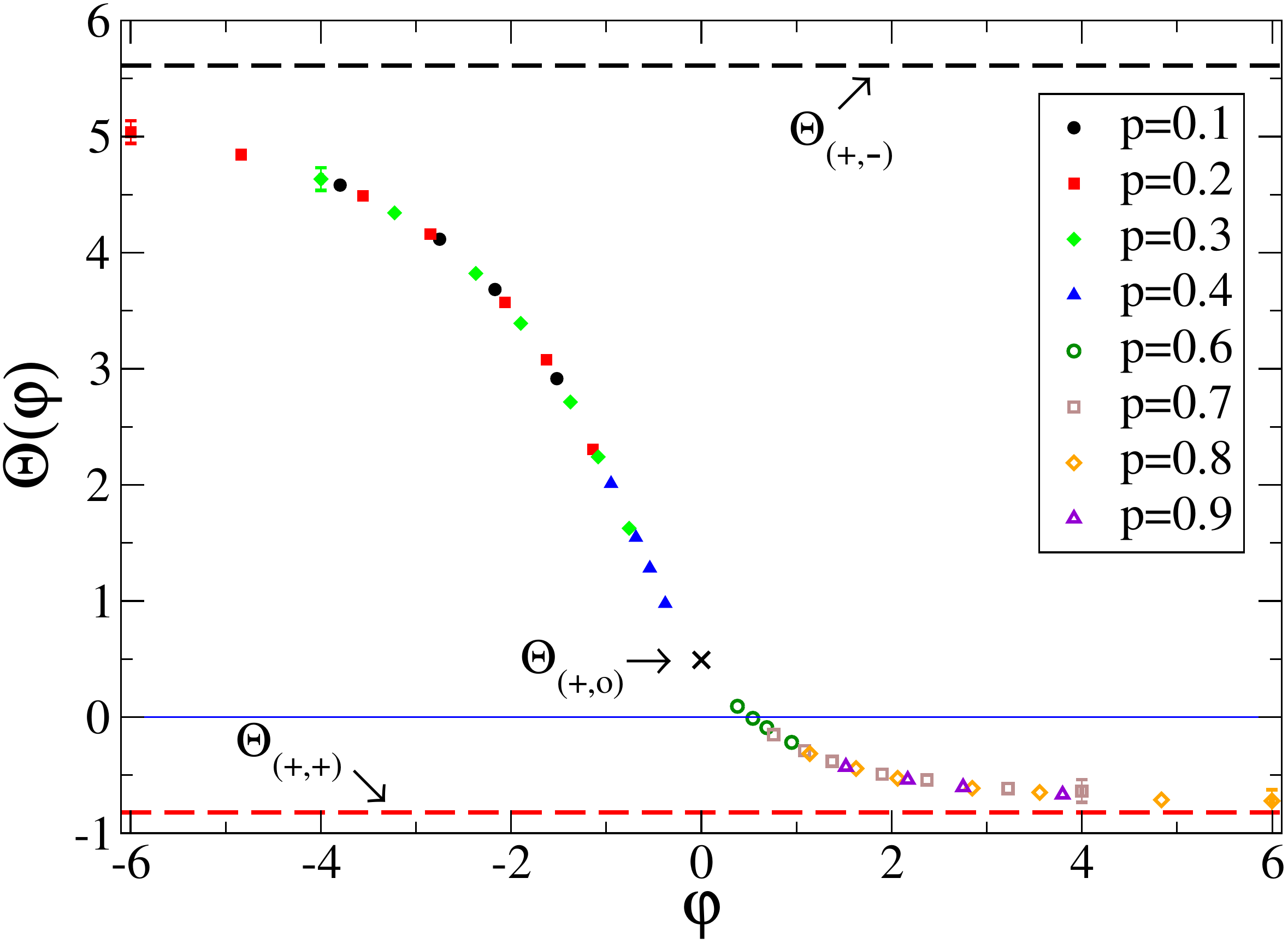}
\caption{(Color online) Crossover scaling function $\Theta(\varphi=(p-1/2)(L-1/2+c)^y)$ for $p=0.1$, $0.2$, $0.3$, $0.4$, $0.6$, $0.7$, $0.8$, $0.9$, and lattice sizes $L=8$, $12$, $16$, $24$, $32$, $48$, $64$. The results for $\Theta(\varphi)$ have been obtained by inverting Eq.~(\ref{casimir_fss_crossover}), using $F_{\rm bulk}(\beta_c)=-0.0757368(4)$ \cite{Hasenbusch-10c}, $c=-1.2(1)$, and $y=y_{h_1}=0.7249(6)$ \cite{Hasenbusch-11}. The curve $\Theta(\varphi)$ saturates at $\Theta(\varphi\rightarrow -\infty)=\Theta_{(+,-)}=5.613(20)$~\cite{Hasenbusch-10c} and $\Theta(\varphi\rightarrow +\infty)=\Theta_{(+,+)}=-0.820(15)$ \cite{Hasenbusch-10c}, which are indicated by dashed lines (see Sec.~\ref{sec:crossover}). At $\varphi=0$ the scaling function $\Theta(\varphi)$ attains the value $\Theta(\varphi=0)=\Theta_{(+,o)}=0.492(5)$ \cite{PTTD-13}, which is indicated by a cross (see Sec.~\ref{sec:phalf}).
The points with displayed error bars correspond to the film thickness $L=64$; for these data the final accuracy is mainly due to the error bars of $\Delta F$ and of $F_{\rm bulk}(\beta_c)$. The omitted error bars are smaller than, or of the size of, the symbols.}
\label{theta_crossover}
\end{figure}
In order to further strengthen this hypothesis, we have reanalyzed the data for $\Delta F(\beta=\beta_c=0.387721735,L,L_\parallel,p)$ with $p\ne 1/2$. By inserting Eq.~(\ref{crossover}) in Eq.~(\ref{force_DeltaF}) and assuming corrections to scaling analog to Eq.~(\ref{casimir_fss}), we find
\begin{multline}
\Delta F(\beta_c,L,L_\parallel,p)= F_{\rm bulk}(\beta_c)\\
-\frac{1}{(L-1/2+c)^3}\Theta\left(\varphi=(p-1/2)(L-1/2+c)^y\right),
\label{casimir_fss_crossover}
\end{multline}
where we have defined $\Theta(\varphi)\equiv\theta(\tau=0,\varphi)$ as the scaling function, which describes the crossover behavior at criticality. It can be obtained by inverting Eq.~(\ref{casimir_fss_crossover}); to this end, we use $F_{\rm bulk}(\beta_c)=-0.0757368(4)$ \cite{Hasenbusch-10c} and $c=-1.2(1)$, as inferred from Table \ref{fit_results_phalf}. In Fig. \ref{theta_crossover} we show $\Theta(\varphi)$ as a function of the scaling variable $\varphi=(p-1/2)(L-1/2+c)^y$, computed assuming the identification of the exponent $y=y_{h_1}=0.7249(6)$ \cite{Hasenbusch-11}. The results shown in Fig.~\ref{theta_crossover} have been obtained using the data of $\Delta F(\beta=\beta_c=0.387721735,L,L_\parallel,p)$ for the values of $p$ considered above, supplemented by some additional MC data with $p=0.1$, $0.4$, $0.6$, $0.9$ for $L=8$, $12$, $16$, and $24$. Within the statistical accuracy we observe a scaling collapse for all the lattice sizes and values of $p$, confirming the validity of Eq.~(\ref{crossover}) and the identification of the exponent $y$ with the RG dimension of the surface field at the ordinary UC.

\section{Summary}
\label{sec:summary}
We have numerically investigated the critical Casimir force in a film geometry, where one surface exhibits a homogeneous adsorption preference, and the opposing surface displays a random local adsorption preference, characterized by a parameter $p$, which measures, on average, the portion of the surface that prefers one component. When $p=1/2$, on average there is no preferential adsorption for one component and the resulting critical Casimir force belongs to the $(+,o)$ UC, supporting the validity of a Harris-type criterion \cite{DN-90} for uncorrelated random surface field, in the limit of infinite amplitude. Thus, for $p=1/2$ a disordered substrate allows to effectively realize Dirichlet {\bc}, which generically do not hold for fluids; Dirichlet {\bc} are also obtained in the presence of a chemically striped surface, in the limit of small stripes width \cite{PTTD-13}. A concurrent study considered the critical Casimir force in the presence of random Gaussian fields with vanishing mean value on both confining surfaces \cite{MVDD-15}, confirming the stability of the ordinary UC against weak disorder, but reporting a different scaling behavior for infinitely strong disorder.

In the present case, when $p>1/2$ (respectively, $p<1/2$) the critical Casimir force belongs to the $(+,+)$ [respectively, $(+,-)$] UC. We observe significant crossover effects and provide evidence that the perturbation associated with the deviation from the $p=1/2$ fixed point can be identified with the surface field.

The present setup can be experimentally realized by monitoring the thickness of a wetting layer of a binary liquid mixture close to criticality on a disordered substrate. Another possibility is provided by considering a spherical colloidal particle in front of such a disordered substrate, provided that the radius of the particle is much larger than its distance to the wall.

\section*{Acknowledgments}
The author is grateful to H.~W.~Diehl and S.~Dietrich for useful discussions, to M.~Hasenbusch and O.~Vasilyev for providing the MC data for comparison, and to A.~Macio\l ek for useful communications. The author acknowledges support from the Max Planck Institute for the Physics of Complex Systems in Dresden, Germany, where this work began.

\bibliographystyle{apsrev4-1_custom}
\bibliography{francesco}

\end{document}